# Inertia in 4D-mechanics

Yurii A. Spirichev

The State Atomic Energy Corporation ROSATOM, "Research and Design Institute of Radio-Electronic Engineering" - branch of Federal Scientific-Production Center "Production Association "Start" named after Michael V.Protsenko", Zarechny, Penza region, Russia

E-mail: yurii.spirichev@mail.ru

05.06.2020

**Abstract**

An important methodological problem of theoretical mechanics related to inertia is discussed. Analysis Inertia is performed in four-dimensional Minkowski space-time based on the law of conservation of energy-momentum. This approach allows us to combine the laws of conservation of momentum and angular momentum into a single law and separate the forces of inertia that actually exist in nature from the imaginary forces introduced to simplify calculations or arising from the transition from one frame of reference to another. From the energy-momentum tensor, in a non-relativistic approximation, the equation of balance of inertia forces existing in nature for a moving continuous medium and a material point in an inertial frame of reference is obtained. It follows from this equation that the pseudo-Euclidean geometry of our world plays an important role in the manifestation of inertia forces. The tensor and the balance equations of all inertia forces in a continuous medium a moving with angular acceleration are obtained. This allows uniform formulation and presentation of the original categories and concepts of classical and relativistic mechanics in educational and scientific literature within the framework of the current paradigm.

**1 Introduction**

The foundations of theoretical mechanics were laid by Newton in his work "Mathematical principles of natural philosophy" [1], where he gave the formulation of the laws of mechanics of a material point. Many of Newton's contemporaries criticized his second law, related to the force of inertia. The criticism was based on the fact that the source of this force was not clear, i.e. there was no way to tell where it came from. This led to the fact that some physicists began to call the forces of inertia fictitious forces. Newton clearly understand the peculiarity of the force of inertia and called it the innate force of matter: «The innate power of matter is its inherent power of resistance, by which every single body, so far as it is left to itself, maintains its state of rest or uniform rectilinear motion…Therefore the innate force could be very intelligibly called the force of inertia… The manifestation of this force can be considered in two ways-both as resistance and as pressure». On this issue, L. Euler wrote «The manifestation of inertia is different from that of ordinary forces. Therefore, to avoid any confusion on this ground, we will omit the word "force" and call the property of bodies in question simply inertia» [2]. A. Yu. Ishlinsky noted that no new concept in mechanics brought so much trouble and confusion in the future, as the Newtonian force of inertia [3]. To solve this problem, a significant contribution was made by E. Mach, who proposed the principle that inertia forces arise and act on bodies when they are accelerated relative to all distant masses of the Universe [4]. Formulated by A. Einstein, the equivalence principle [5], which linked the forces of inertia with gravitational forces and experimentally confirmed with an accuracy of $10^{-14}$ [6], indicates the nature of the forces of inertia. Modern works [7-13] are also devoted to attempts to solve this problem.

When considering non-inertial reference systems and transitions from one reference system to another, especially if they move relative to each other with dynamically changing acceleration, the concepts of forces acting in them still cause numerous disputes and misunderstandings. As a result, the terms "centrifugal force of inertia", "centrifugal force", "Coriolis force", "centripetal force", "Dalembert force of inertia", "Euler force of inertia", "Kepler force of inertia", "portable force of inertia", etc. are used in mechanics. which are not only understood in different ways by different authors, but are sometimes defined mathematically in different ways.

All controversial opinions related to the effects of inertia can be divided into two parts. The first part concerns the question whether the effects of inertia are forces (according to Newton) or they are not forces (according to Euler). The second part concerns the question of which forces or effects of inertia actually exist in nature, and which are imaginary. After solving the question of the second part and determining the actual effects or forces of inertia, the question of the first part passes into the category of terminological questions, when to call the real effects of inertia "forces of inertia", just "forces" or "inertia" is the taste of the researcher. To solve the discussion issues of the second part it is necessary to find a criterion for determining the actual forces or effects of inertia. Forces are derived from energy, and therefore the consideration of inertia forces based on the law of conservation of energy allows us to determine their reality or fictitiousness.

In theoretical mechanics, separate laws of conservation of energy, momentum, and angular momentum are formulated. However, the change in views of space and time as a single continuum that occurred in the early twentieth century in the works of A. Poincare, G. Minkowski, H. Lorenz and A. Einstein leads to a single law of conservation of energy-momentum, described by a single system of equations arising from the energy-momentum tensor. Therefore, the revision of views on space-time created the prerequisites for a methodically more complete description of the laws of mechanics, which allows us to solve the second part of the discussion on the forces of inertia. According to Einstein's General theory of relativity, space-time is considered pseudo-Riemannian, but in the small vicinity of a local point, in the absence of a strong gravitational field, it can be considered a linear Minkowski space. In this regard, in this paper, the law of conservation of energy-momentum and inertia will be considered in the Minkowski space, i.e. from the point of view of special theory of relativity (STR), considering it experimentally confirmed with high accuracy.

The purpose of this article is to determine, based on the law of conservation of energy-momentum, the real effects or forces of inertia acting on a moving continuous medium and a material point in Minkowski space.

In this paper, four-dimensional vector quantities have a representation with imaginary spatial components and a real time component [14]. With this representation of 4-vectors, the geometry of Minkowski space-time corresponds to the four-dimensional geometry of Euclid in a homogeneous space. All movements of the medium and the material point will be considered in the non-relativistic approximation, taking the relativistic coefficient $1/\sqrt{1-v^2/c^2}$ for low speeds of movement equal to one. In relativistic mechanics, a four-dimensional radius vector has a representation $\mathbf{R}_\nu(c \cdot t, i \cdot \mathbf{r})$, where c - is the speed of light, t - is time, and $\mathbf{r}$ - is a three-dimensional radius-vector [15]. In accordance with this representation, the four-dimensional velocity of a continuous medium or material point, for uniformity of the dimensions of quantities in the three – dimensional representation, will be described as a 4-vector $\mathbf{V}_\nu(c, i \cdot \mathbf{v})$, where $\mathbf{v}$ - is the three-dimensional velocity of the medium. We will describe the four-dimensional density of the mechanical momentum of a continuous medium in the form of a 4-vector $\mathbf{P}_\nu(c \cdot m, i \cdot \mathbf{p})$, where m and $\mathbf{p}$ are, respectively, the mass density and the three-dimensional vector of the momentum density of the medium particles. For the accepted representation of the space-time geometry, the four-dimensional partial derivative operator has the form $\partial_\mu(\partial_t/c, i \cdot \nabla)$. For the accepted description of this operator and four-dimensional vector quantities, it is possible not to distinguish covariant and contravariant indices.

**2 Tensor, equations of the law of conservation of energy-momentum and inertia force in a continuous medium**

We find the energy-momentum tensor (EMT) of the medium in the form of the tensor product of 4-velocity $\mathbf{V}_\mu(c, i \cdot v)$ by 4-momentum $\mathbf{P}_\nu(c \cdot m, i \cdot \mathbf{p})$ [16]:

$$T_{\mu\nu} = \mathbf{V}_\mu \otimes \mathbf{P}_\nu = \begin{pmatrix} m \cdot c^2 & i \cdot c \cdot p_x & i \cdot c \cdot p_y & i \cdot c \cdot p_z \\ i \cdot c \cdot p_x & -v_x \cdot p_x & -v_x \cdot p_y & -v_x \cdot p_z \\ i \cdot c \cdot p_y & -v_y \cdot p_x & -v_y \cdot p_y & -v_y \cdot p_z \\ i \cdot c \cdot p_z & -v_z \cdot p_x & -v_z \cdot p_y & -v_z \cdot p_z \end{pmatrix} \quad (1)$$

The energy-momentum density tensor (1) is symmetric. Its trace is a well-known [15] 4-invariant of the mechanical energy density $I = \delta_{\mu\nu} T_{\mu\nu} = m \cdot c^2 - \mathbf{p} \cdot \mathbf{v}$. The equation of the law of conservation of energy-momentum for a closed physical system has the form 4-divergence 4-energy-momentum tensor $\partial_\mu T_{\mu\nu} = 0$. Let's write this equation in expanded form as a system of equations:

$$\partial_t mc + c\nabla \cdot \mathbf{p} = 0 \quad \text{or} \quad \partial_t m + \nabla \cdot \mathbf{p} = 0 \quad (2)$$

$$2\partial_t \mathbf{p} + \mathbf{p} \cdot (\nabla \cdot \mathbf{v}) + (\mathbf{v} \cdot \nabla)\mathbf{p} + (\mathbf{p} \cdot \nabla)\mathbf{v} + \mathbf{v} \cdot (\nabla \cdot \mathbf{p}) = 0 \quad \text{or}$$

$$2\partial_t \mathbf{p} + \nabla(\mathbf{p} \cdot \mathbf{v}) + \mathbf{p} \cdot (\nabla \cdot \mathbf{v}) + \mathbf{v} \cdot (\nabla \cdot \mathbf{p}) - \mathbf{p} \times \nabla \times \mathbf{v} - \mathbf{v} \times \nabla \times \mathbf{p} = 0 \quad (3)$$

Eq. (2) describes the law of conservation of mass density (energy), and Eq. (3) describes the law of conservation of momentum density. Eq. (3) includes a description of the rotational movements of the medium (the fifth and sixth terms), i.e. it also includes a description of the law of conservation of momentum density. Consequently, in four-dimensional Minkowski space, the three-dimensional laws of conservation of momentum and angular momentum merge into a single law of conservation of momentum. Replacing Eq. (3) $\nabla \cdot \mathbf{p}$ in $-\partial_t m$ the fourth term with Eq. (2), we get the equation:

$$m \cdot \partial_t \mathbf{v} + [\mathbf{v} \cdot \partial_t m + \mathbf{p} \cdot (\nabla \cdot \mathbf{v}) + \nabla(m \cdot \mathbf{v}^2) - \mathbf{p} \times \nabla \times \mathbf{v} - \mathbf{v} \times \nabla \times \mathbf{p}]/2 = 0 \quad (4)$$

Eq. (3) is an equation for the balance of the density of all inertia forces acting in an ideal continuous elastic medium in the absence of external forces. The first term of Eq. (3) is the density of Newton's inertial force. The forces described by the second, third, and fourth terms depend on the rate of change in the density of the medium. For a medium velocity constant in time, Newton's inertia force is equal to zero, and for this case Eq. (3) can be written as:

$$\nabla(m \cdot \mathbf{v}^2) + m \cdot \mathbf{v}(\nabla \cdot \mathbf{v}) = \mathbf{v} \times \nabla \times (m \cdot \mathbf{v}) + m \cdot \mathbf{v} \times \nabla \times \mathbf{v} \qquad (5)$$

The right side of the equation describes the density of the centrifugal forces of inertia of the rotation of the medium, since they are directed away from the axis of rotation. It follows from this equation that the forces on its left side compensate for the centrifugal forces on the right side, therefore, the forces on the left side of Eq. (5) can be considered centripetal forces. Thus, the continuous medium in the Minkowski space is affected by six types of inertial forces included in Eq. (3). Any other forces of inertia are fictitious or imaginary, since they are not included in the equation of the conservation law energy-momentum (3).

**3 Tensor, equations of the law of conservation of energy-momentum and inertia force for a material point**

The description of the law of conservation of energy-momentum for a material point follows from the EMT (1), the system of Eqs. (2) and (3) for a continuous medium, and the condition of constancy in time and space of the mass density of the considered particle of the continuous medium. Denoting the constant mass of a material particle by $m_0$, we obtain the tensor and the energy-momentum conservation equations in the form:

$$T_{\mu\nu} = m_0 \cdot \mathbf{V}_\mu \otimes \mathbf{V}_\nu = m_0 \cdot \begin{pmatrix} c^2 & i \cdot c \cdot v_x & i \cdot c \cdot v_y & i \cdot c \cdot v_z \\ i \cdot c \cdot v_x & -v_x \cdot v_x & -v_x \cdot v_y & -v_x \cdot v_z \\ i \cdot c \cdot v_y & -v_y \cdot v_x & -v_y \cdot v_y & -v_y \cdot v_z \\ i \cdot c \cdot v_z & -v_z \cdot v_x & -v_z \cdot v_y & -v_z \cdot v_z \end{pmatrix}$$

$$\nabla \cdot \mathbf{v} = 0 \qquad (6)$$

$$m_0 \partial_t \mathbf{v} + m_0 \nabla(\mathbf{v}^2) + 2m_0 \mathbf{v}(\nabla \cdot \mathbf{v}) - 2m_0 \mathbf{v} \times \nabla \times \mathbf{v} = 0 \qquad (7)$$

Taking into account Eq. (6), Eq. (7) can be written in the form:

$$m_0 \partial_t \mathbf{v} + m_0 \nabla \mathbf{v}^2 - 2m_0 \mathbf{v} \times \nabla \times \mathbf{v} = 0 \qquad (8)$$

This equation is an equation of the inertial motion of a material point or an equation for the balance of inertial forces acting on a material point in Minkowski space. The first term of Eq. (8) describes Newton's force. The second term describes the centripetal force and the third term describes the centrifugal force. Thus, it follows from the law of conservation of energy-momentum that, in the general case, three types of inertial forces act on a material point moving with acceleration in Minkowski space. There are no other inertial forces for the case under consideration, since they contradict the energy-momentum conservation law. All other forces introduced to simplify calculations or resulting from the transition from one frame of reference or coordinate system to other systems are fictitious or imaginary.

If an external four-dimensional force $F_\mu$ ($\mu=0,1,2,3$) acts on a material point, then the force balance Eqs. (6) and (7) can be written as:

$$c \cdot m_0 \cdot \nabla \cdot \mathbf{v} = F_0 \qquad (9)$$

$$m_0 \cdot \partial_t \mathbf{v} + m_0 \cdot \nabla(\mathbf{v}^2)/2 + m_0 \cdot \mathbf{v} \cdot (\nabla \cdot \mathbf{v}) - m_0 \cdot \mathbf{v} \times \nabla \times \mathbf{v} = \mathbf{F}_{123} \qquad (10)$$

This system of equations can be considered as a generalization of Newton's second law for a material point moving with a nonrelativistic velocity in Minkowski space under the action of a four-dimensional force $F_\mu$.

In mechanics, the question of the possibility of the so-called "unsupported motion" was discussed for a long time, i.e. accelerated motion of a closed physical system only due to the action of internal forces, without the rejection of mass. This effect was found V. N. Tolchin in experiments with rotating flywheels [17]. To date, no satisfactory explanation for this effect has been found. According to the existing ideas in classical mechanics, such a movement is impossible, since it contradicts the laws of conservation of momentum and angular momentum. This conclusion follows from the fact that in classical mechanics the laws of conservation of momentum and angular momentum are considered as two separate and independent laws [3]. Eq. (8), which combines the laws of conservation of momentum and angular momentum into a single, interconnected law of conservation of momentum, shows the possibility of such a movement. It follows from this equation that in a closed physical system it is possible to obtain a change in momentum by changing the angular momentum. Thus, Eq. (8) makes it possible to theoretically explain the Tolchin effect on the basis of the pseudo-Euclidean geometry of SRT.

If we take the rest energy $m_0 c^2$ from the EMT, then the dimensionless tensor will be its factor:

$$T_{\mu\nu} = \mathbf{V}_\mu \otimes \mathbf{P}_\nu = m_0 c^2 \cdot \begin{pmatrix} 1 & i \cdot v_x/c & i \cdot v_y/c & i \cdot v_y/c \\ i \cdot v_x/c & -v_x^2/c^2 & -v_x \cdot p_y & -v_x \cdot v_z/c^2 \\ i \cdot v_y/c & -v_y \cdot v_x/c^2 & -v_y^2/c^2 & -v_y \cdot v_z/c^2 \\ i \cdot v_z c & -v_z \cdot v_x/c^2 & -v_z \cdot v_y/c^2 & -v_z^2/c^2 \end{pmatrix} \qquad (11)$$

The diagonal components of this dimensionless tensor have the Minkowski space metric signature (1, -1, -1, -1), and all dimensionless components except for $T_{00}$ depend on the velocity of the material point. This can be seen as a curvature of the geometry of the Minkowski space depending on the speed of matter. From this we can conclude that the pseudo-Euclidean geometry of our world plays an important role in the manifestation of the forces of inertia. At low speeds, the curvature of the geometry will be small. At relativistic velocities, taking into account the relativistic coefficient, the dimensionless spatial coefficients of the metric and the corresponding space curvature can be significant. This is natural, since as the speed increases, so does the kinetic energy. Thus, already within the framework of SRT, the idea of the curvature of the space-time geometry follows from the EMT.

**4 Tensor and equations of balance of inertia forces with angular acceleration**

In the equations of conservation of energy-momentum (2), (3), (6), (7) for a medium and a material point, the inertial forces associated with curvilinear motion depend on the speed of motion, and not on acceleration in time. To obtain the equations for the balance of inertial forces that arise when moving with angular acceleration, it is necessary to find the four-dimensional derivative of the EMT. This derivative is the tensor of inertia forces of the third rank:

$$F_{\lambda\mu\nu} = \partial_\lambda T_{\mu\nu} = \partial_\lambda (\mathbf{V}_\mu \otimes \mathbf{P}_\nu) = \partial_\lambda \begin{pmatrix} m \cdot c^2 & i \cdot c \cdot p_x & i \cdot c \cdot p_y & i \cdot c \cdot p_z \\ i \cdot c \cdot p_x & -v_x \cdot p_x & -v_x \cdot p_y & -v_x \cdot p_z \\ i \cdot c \cdot p_y & -v_y \cdot p_x & -v_y \cdot p_y & -v_y \cdot p_z \\ i \cdot c \cdot p_z & -v_z \cdot p_x & -v_z \cdot p_y & -v_z \cdot p_z \end{pmatrix} \quad (12)$$

After differentiation, we get:

$$F_{\lambda\mu\nu} = \partial_\lambda T_{\mu\nu} = \partial_\lambda (\mathbf{V}_\mu \otimes \mathbf{P}_\nu) = (\partial_\lambda \mathbf{V}_\mu) \otimes \mathbf{P}_\nu + \mathbf{V}_\mu \otimes (\partial_\lambda \mathbf{P}_\nu) \quad (13)$$

From the tensor of inertia forces (12) in the form of its divergences follows the system of equations for the balance of inertial forces:

$$F_{\lambda\mu\nu} = \partial_\lambda T_{\mu\nu} = \partial_\lambda (\mathbf{V}_\mu \otimes \mathbf{P}_\nu) = m_0 \cdot \partial_\lambda (\mathbf{V}_\mu \otimes \mathbf{V}_\nu) = m_0 \cdot A_{\lambda\mu\nu} \quad (14)$$

For a material point, from equation (10) one can obtain an equation of accelerated rotational motion describing gyroscopic effects. To do this, take the rotor from its both parts:

$$m_0 \cdot \partial_t \mathbf{\Omega} - m_0 \cdot \nabla \times \mathbf{v} \times \mathbf{\Omega} = \nabla \times \mathbf{F}_{123} \quad (15)$$

Where is the angular velocity of the material point. When the right side of Eq. (15) is equal to zero, it describes the accelerated inertial rotation observed in the Dzhanibekov effect in weightlessness.

**5 Conclusion**

The revision of views on space-time created the prerequisites for a methodically more complete description of the general laws of mechanics. Analysis of the forces of inertia in the pseudo-Euclidean Minkowski space on the basis of the law of conservation of energy-momentum made it possible to separate the forces of inertia that actually exist in nature from imaginary forces introduced to simplify calculations or that arise when switching to another frame of reference.

From the EMT in the nonrelativistic approximation, a system of two equations is obtained that describes the laws of conservation of energy, momentum, and angular momentum in a continuous medium. The second equation of this system can be considered as the equation of balance of actually existing inertial forces for a continuous medium and a material point moving in an inertial frame of reference. It follows from the equation of inertial motion of a material point that in the Minkowski space it is affected by Newton's force, centripetal force and centrifugal force, which can be considered as really acting forces that ensure the fulfillment of the energy-momentum conservation law. All other forces introduced to simplify calculations or arising from the transition from one frame of reference to another are imaginary or fictitious, since they are not included in the energy-momentum conservation law.

It follows from the energy-momentum conservation equation that inertia forces arise during any movement of a material point. In rectilinear motion at a constant speed, Newton's force is zero, and the centripetal and centrifugal forces have different signs and cancel each other out. From the momentum conservation equation, one can exclude the mass of a material point and obtain a kinematic equation of motion independent of the mass. It follows that, speaking in the language of Newton, the pseudo-Euclidean geometry of our world plays an important role in the manifestation of the innate force of matter.

A tensor of the third rank and equations for the balance of inertia forces in the motion of a medium or a material point with angular acceleration are obtained.

It is shown that in the framework of SRT it is possible to represent the curvature of the space-time geometry depending on the speed of matter.

The analysis of the forces of inertia in the Minkowski space and the conclusions drawn from it make it possible to uniformly formulate and present the initial categories and concepts of classical and relativistic mechanics in educational and scientific literature.